\documentclass[12pt,noshowpacs,nofootinbib,notitlepage,amsmath,amssymb]{revtex4-2}
\usepackage{setspace}
\usepackage[top=1in,bottom=1in,left=1in,right=1in]{geometry}
\usepackage{graphicx,color}
\usepackage[colorlinks=true,citecolor=blue,linkcolor=blue,urlcolor=blue]{hyperref}
\usepackage{enumerate}
\usepackage{array}
\usepackage{bm}
\usepackage{xcolor}

\begin{document}
\title{\color{blue}\Large Superselected ghost theory: real spectrum}

\author{Bob Holdom}
\email{bob.holdom@utoronto.ca}
\affiliation{Department of Physics, University of Toronto\\ Toronto, Ontario, Canada  M5S 1A7}

\begin{abstract}
Quantum field theories with ghosts can be unitary and perturbatively stable, yet the negative-norm states of their conserved indefinite inner product obstruct a probabilistic interpretation. This problem is especially relevant to renormalizable quantum gravity. The focus of this paper is the real-spectrum regime, in which the interacting Hamiltonian has an exact $\mathbb{Z}_2$ symmetry $Q$, called exact ghost parity. Its eigenvalue on each energy eigenstate equals the sign of the norm, and it reduces to free ghost parity at zero coupling. Imposing $Q$ as a superselection charge defines a new theory in which the physical states have definite ghost parity and the observables commute with $Q$. The native Born rule then yields non-negative probabilities, while the optical theorem acquires a direct probabilistic interpretation. The propagator decomposes into $Q$-sector spectral representations with fixed-sign spectral functions and no complex poles on the physical sheet. Finally, a similarity transformation yields a Hermitian perturbation theory that preserves free ghost parity order by order, making the exact superselection structure perturbatively manifest.
\end{abstract}

\maketitle

\section{Introduction}\label{s1}

We refer to quantum field theories with at least one wrong-sign kinetic term as ghost theories. Such terms arise in four-derivative theories, which are of particular interest to quantum gravity. Quantum quadratic gravity, for example, is perturbatively renormalizable~\cite{Stelle:1976gc,Salvio:2018crh}. Nevertheless, this approach is commonly set aside because four-derivative theories are often regarded as fundamentally problematic due to instability, non-unitarity, or the absence of a probability interpretation. However, with a suitable quantization, the ghost theories considered here are unitary and perturbatively stable. The central difficulty is, instead, that the native inner product is indefinite. The question is therefore how to extract probabilities from a theory whose time evolution is well defined but whose natural inner product is not positive definite.

In a ghost theory, a ghost parity operator $Q$ exists that commutes with the full interacting Hamiltonian. However, it is not a symmetry of the free and interaction Hamiltonians separately. Consequently, it is not visible order by order in ordinary perturbation theory and need not be evident in the defining Lagrangian. This distinguishes it from \emph{free} ghost parity, which is defined by the native indefinite inner product and labels free states according to the signs of their norms. The free ghost parity commutes with the free Hamiltonian but not with the interacting one. Exact ghost parity, by contrast, commutes with the full Hamiltonian and reduces to free ghost parity in the zero-coupling limit.

Such an exact symmetry is familiar from pseudo-Hermitian \cite{Mostafazadeh:2001jk} and PT-symmetric \cite{Bender:2002vv} quantum mechanics. There, the role of $Q$ is played by the operator $\mathcal{C}$, which is constructed from energy eigenstates and is generally nonlocal~\cite{Mostafazadeh:2003gz,Bender:2007nj}. The Hamiltonian is self-adjoint with respect to both the native inner product and a modified inner product constructed using $\mathcal{C}$. When the modified inner product is positive definite, the spectrum is real \cite{Mostafazadeh:2001nr,Mostafazadeh:2002id}.

The treatment of 0+1D ghost theory in \cite{Holdom:2024onr} used this positive-definite inner product. The native inner product defined the theory itself, while the positive-definite inner product was used to extract probabilities through a modified Born rule. In this paper, we follow a different route that does not require modifying the Born rule.

To begin, $\mathbb{Z}_2$ symmetries of the Hamiltonian are straightforward to construct once the energy eigenstates are known: one simply assigns $+1$ or $-1$ to each projector in the spectral decomposition of the identity. There is therefore an infinite family of such operators. The exact ghost parity is one member of this family, and its mere existence does not affect the theory. In the standard treatment of ghost theories, superpositions of states with opposite ghost parity are still considered, and physical observables are not required to commute with $Q$. Moreover, one still uses ordinary perturbation theory, which explicitly violates the ghost parity symmetry of the exact Hamiltonian.

Using $Q$ to define a superselection rule changes this situation. It defines a new theory, a superselected ghost theory, which generally differs from the original ghost theory. The theory can lie in either of two regimes, depending on whether its spectrum is real. Which regime occurs is a dynamical question. Studies of PT-symmetric quantum mechanics have identified many non-Hermitian Hamiltonians with real spectra, and the same phenomenon occurs in 0+1D ghost theories \cite{Holdom:2024onr}. These results motivate our focus on the real-spectrum regime. An accompanying paper considers the effect of superselection when the spectrum contains complex-conjugate pairs of energies \cite{entang}.

Each energy eigenstate has an exact ghost parity that matches the sign of its norm. Because physical superpositions must lie within a single sector, every physical state has a definite norm sign. We assume that the exact vacuum is unique and continuously connected to the free Fock vacuum, so that it has $Q = +1$ and a positive norm. Physical observables are operators that commute with $Q$.

Without the superselection rule, the Born rule, which is inversely proportional to the norms of the initial and final states, produces negative probabilities for transitions between states of opposite norm sign. A superselected ghost theory separates the two sectors by excluding physical superpositions across them and forbidding transitions between them. The native Born rule then produces only positive probabilities.

The superselection rule has a similar effect on the optical theorem. Perturbation theory computes the left-hand side, the imaginary part of a forward scattering amplitude, using the native inner product. This result agrees with the right-hand side when the latter is also formulated using the native inner product, thereby establishing unitarity. However, the right-hand side has a probabilistic interpretation only when calculated using the positive inner product. Once the superselection rule is imposed, the native inner product also has the correct probabilistic content, making the positive inner product unnecessary in this context.

The superselection rule also decomposes the spectral representation of propagators into two independent contributions, one from each exact ghost parity sector. Each contribution has its own standard analytic structure. This decomposition prevents complex-conjugate poles from appearing on the physical sheet of any propagator. It is a nontrivial consistency condition for the real-spectrum regime that would not hold without the superselection rule.

In Section~\ref{s2}, we review the matrix notation introduced in \cite{Holdom:2024onr}, which clarifies the effects of the native and positive inner products, and extend it to 3+1D QFT. Sections \ref{s3} and \ref{s4} discuss the optical theorem and the spectral representation, respectively. We also discuss the differences caused by the choice of inner product. Although $Q$ is nonlocal, these applications rely only on its action on exact energy eigenstates or on the asymptotic Fock space, where $Q$ is diagonal.

The original perturbation theory of the ghost theory, when summed to all orders, is expected to be consistent with the exact ghost symmetry of the Hamiltonian. However, neither the ghost symmetry nor the superselection rule is respected at any finite order. In Section~\ref{s5}, we introduce a similarity-transformed perturbation theory that preserves ghost parity order by order, thus becoming consistent with the superselection rule. An accompanying paper \cite{pertur} develops this construction further and shows how the nonlocality of $Q$ is handled in perturbation theory. This construction yields the effective Hamiltonian for the superselected theory, where new nonlocal interactions at second order and higher are determined by the theory. In effect it is a modified form of old-fashioned perturbation theory and so is local in time but is spatially nonlocal.

\section{From 0+1D to 3+1D QFT}\label{s2}

\subsection{0+1D: Pseudo-Hermitian quantum mechanics}

We begin with 0+1D QFT, where field-theoretic constructions such as the full Feynman propagator already apply~\cite{Holdom:2024onr}. In the occupation-number basis, the Hamiltonian is represented by a Hermitian matrix $\mathbf{H}$. Quantization with a positive spectrum for the free Hamiltonian gives the ghost theory its native inner product,
\begin{equation}
\langle\chi|\psi\rangle_\eta = \boldsymbol{\chi}^\dagger\bm\eta\boldsymbol{\psi},
\qquad \bm\eta^2 = \mathbf{1},
\end{equation}
where $\boldsymbol{\chi}$ and $\boldsymbol{\psi}$ are column-vector representations of states. Because $\bm\eta$ has positive and negative eigenvalues, the inner product is indefinite, and the state space is a Krein space rather than a Hilbert space. Time evolution is generated by $\tilde{\mathbf{H}} \equiv \bm\eta\mathbf{H}$, which is $\eta$-self-adjoint:
$\tilde{\mathbf{H}}^\dagger\bm\eta = \bm\eta\tilde{\mathbf{H}}$. Equivalently, the native inner product is preserved under time evolution,
\begin{equation}
\frac{d}{dt}\langle\chi|\psi\rangle_\eta
= i\boldsymbol{\chi}^\dagger(\tilde{\mathbf{H}}^\dagger\bm\eta
  - \bm\eta\tilde{\mathbf{H}})\boldsymbol{\psi} = 0.
\end{equation}

For a single ghost degree of freedom with occupation-number basis $|n\rangle$, the state vector has components $(\boldsymbol{\psi})_n = \psi_n$, and the matrix representing an operator $A$ is defined by $(\mathbf{A})_{mn} = \langle m|A|n\rangle$. In particular,
$\langle m|n\rangle_\eta = (\bm\eta)_{mn} = (-1)^n\delta_{mn}$. These signs enter the completeness relation,
\begin{equation}
\mathbf{1}= \sum_n \frac{|n\rangle\langle n|}{\langle n|n\rangle_\eta},
\end{equation}
as well as $\langle n|\psi\rangle_\eta = (-1)^n\psi_n$. The column-vector representation of $A|\psi\rangle$ is therefore
$\tilde{\mathbf{A}}\boldsymbol{\psi} \equiv \bm\eta\mathbf{A}\boldsymbol{\psi}$. Similarly, $H|\psi\rangle$ is represented by
$\bm\eta\mathbf{H}\boldsymbol{\psi} = \tilde{\mathbf{H}}\boldsymbol{\psi}$~\cite{Holdom:2024onr}.
Taking the inner product with $\langle\chi|$ gives
$\langle\chi|A|\psi\rangle = \boldsymbol{\chi}^\dagger\mathbf{A}\boldsymbol{\psi}
= (\bm\eta\boldsymbol{\chi})^\dagger\tilde{\mathbf{A}}\boldsymbol{\psi}$.
This notation extends to general ghost theories.

The matrix representing free ghost parity is $\bm\eta$ itself, with
$[\bm\eta,\tilde{\mathbf{H}}_0] = 0$ and $\bm\eta^2 = \mathbf{1}$. The free states are eigenstates of $\bm\eta$, with eigenvalues $\pm 1$ that match the signs of their norms. For the interacting theories of interest, however,
$[\bm\eta,\tilde{\mathbf{H}}] \neq 0$. In the real-spectrum regime, the ghost theory admits a second inner product that is conserved and positive definite. It is defined by a Hermitian matrix $\mathbf{G}$,
\begin{equation}
\langle\psi|\chi\rangle_G = \boldsymbol{\psi}^\dagger\mathbf{G}\boldsymbol{\chi}.
\end{equation}
$\mathbf{Q}$ is then introduced through $\mathbf{G} = \bm\eta\mathbf{Q}$, and the conservation of the $G$ inner product is equivalent to
\begin{equation}
[\mathbf{Q},\tilde{\mathbf{H}}] = 0.
\label{e8}
\end{equation}
An explicit construction of $\mathbf{Q}$ that yields a positive-definite $\mathbf{G}$ and satisfies $\mathbf{Q}^2 = \mathbf{1}$ is
\begin{equation}
\mathbf{Q} = \sum_n
\frac{\boldsymbol{\psi}^{\bar{n}}\boldsymbol{\psi}^{\bar{n}\dagger}\bm\eta}
{|\boldsymbol{\psi}^{\bar{n}\dagger}\bm\eta\boldsymbol{\psi}^{\bar{n}}|},
\label{e9}
\end{equation}
where $\bar{n}$ labels the exact energy eigenstates. The eigenvalue of $\mathbf{Q}$ on each energy eigenstate is its ghost parity,
$\mathbf{Q}\boldsymbol{\psi}^{\bar{n}} = \sigma_n\boldsymbol{\psi}^{\bar{n}}$, which matches the sign of its norm:
$\sigma_n = \mathrm{sgn}(\boldsymbol{\psi}^{\bar{n}\dagger}\bm\eta
\boldsymbol{\psi}^{\bar{n}})$. The energy eigenstates are orthogonal with respect to both inner products. Although
free states are eigenstates of $\bm\eta$, exact energy eigenstates are
generally not eigenstates of $\bm\eta$ or $\mathbf{G}$. We have
$\mathbf{G}\boldsymbol{\psi}^{\bar{n}} = \sigma_n\bm\eta
\boldsymbol{\psi}^{\bar{n}}$ and, since $\mathbf{G}$ is Hermitian,
$\boldsymbol{\psi}^{\bar{n}\dagger}\mathbf{G} = \sigma_n
\boldsymbol{\psi}^{\bar{n}\dagger}\bm\eta$. Thus
$\boldsymbol{\psi}^{\bar{n}\dagger}\mathbf{G}\boldsymbol{\psi}^{\bar{n}}
= \sigma_n\boldsymbol{\psi}^{\bar{n}\dagger}\bm\eta
\boldsymbol{\psi}^{\bar{n}} > 0$.

Equation~(\ref{e8}) ties the construction of the $G$ inner product to the energy-eigenstate basis. This relation is well known in pseudo-Hermitian~\cite{Mostafazadeh:2001jk} and PT-symmetric~\cite{Bender:2002vv} quantum mechanics, although the context differs. Those theories typically begin with a non-Hermitian Hamiltonian and seek an inner product that provides both unitarity and positivity. In the ghost theories considered here, the $\eta$ inner product already provides unitarity and continues to play a central role. We carry the positive inner product along, and it becomes important in Section~\ref{s5}.

\subsection{3+1D: Asymptotic Fock space}

To extend the matrix notation to 3+1D, we assume a finite-volume formulation in which the Hamiltonian is self-adjoint with respect to the native inner product, is diagonalizable, and has a real spectrum with no null eigenstates. We also assume that the continuum limit is well defined. Energy eigenstates play a privileged role in the formalism of 3+1D QFT. When a scattering description exists, stable particles admit asymptotic in- and out-states labeled by their physical masses and momenta. The asymptotic state space is therefore naturally organized as a Fock space built from exact one-particle states. This structure makes the spectral role of energy eigenstates more accessible than in 0+1D, where no scattering interpretation exists and the energy levels must be found by solving the interacting theory.

The asymptotic state space can be equipped with either the native inner product $\eta$ or the positive-definite inner product $G$. The same Fock-space construction therefore has both a Krein-space realization and a Hilbert-space realization. The Fock states are eigenstates of exact ghost parity $Q$, whose eigenvalues match the signs of their norms when the native inner product is used. We let $\mathbf{K} = \bm\eta$ or $\mathbf{G}$, noting that $\mathbf{K}^{-1}=\mathbf{K}$ only when $\mathbf{K}=\bm\eta$. In matrix notation, the completeness relation over the Fock basis is
\begin{equation}
\mathbf{1} = \sum_X \int d\Pi_X \frac{\boldsymbol{\psi}^X \boldsymbol{\psi}^{X\dagger} \mathbf{K}}{\boldsymbol{\psi}^{X\dagger} \mathbf{K} \boldsymbol{\psi}^X},
\label{e3}
\end{equation}
\begin{equation}
d\Pi_X = \prod_{k=1}^{n_X} d^3 p_k.
\label{e22}\end{equation}
Here, $n_X$ is the number of particles in state $X$. The completeness relation is identical term by term for either choice of $\mathbf{K}$, since replacing $\bm\eta$ with $\mathbf{G}$ introduces the same factor of $\sigma_X$ in the numerator and denominator. Although we assume finite-volume regularization to retain matrix notation, we use continuum notation such as~(\ref{e22}) when convenient. Norm factors such as those in (\ref{e3}) carry essential sign information. These factors and the definition of $d\Pi_X$ can accommodate different conventions.

The operators $\tilde{\mathbf{H}}$ and $\mathbf{Q}$ are defined by
\begin{equation}
\tilde{\mathbf{H}} = \sum_X \int d\Pi_X \, E_X \frac{\boldsymbol{\psi}^X \boldsymbol{\psi}^{X\dagger} \bm\eta}{\boldsymbol{\psi}^{X\dagger} \bm\eta \boldsymbol{\psi}^X},
\end{equation}
\begin{equation}
\mathbf{Q} = \sum_X \int d\Pi_X \frac{\boldsymbol{\psi}^X \boldsymbol{\psi}^{X\dagger} \bm\eta}{|\boldsymbol{\psi}^{X\dagger} \bm\eta \boldsymbol{\psi}^X|}.
\end{equation}
Because $[\mathbf{Q}, \tilde{\mathbf{H}}] = 0$, the Hamiltonian $\tilde{\mathbf{H}}$ is self-adjoint with respect to both inner products:
$\tilde{\mathbf{H}}^\dagger \mathbf{K} = \mathbf{K} \tilde{\mathbf{H}}$. Consequently,
\begin{equation}
e^{i\tilde{\mathbf{H}}^\dagger t} \mathbf{K} e^{-i\tilde{\mathbf{H}} t} = \mathbf{K} e^{i\tilde{\mathbf{H}} t} e^{-i\tilde{\mathbf{H}} t} = \mathbf{K}.
\end{equation}

\section{The $S$-matrix}\label{s3}

The matrix form of the $S$-matrix is
\begin{align}
\tilde{\mathbf{S}} &= \lim_{t\to\infty,\, t_0\to-\infty} e^{i\tilde{\mathbf{H}}_{\mathrm{as}}t}\, e^{-i\tilde{\mathbf{H}}(t-t_0)}\, e^{-i\tilde{\mathbf{H}}_{\mathrm{as}}t_0}, \label{e10}\\
\tilde{\mathbf{H}}_{\mathrm{as}} &= \tilde{\mathbf{H}} \oplus (\tilde{\mathbf{H}}\otimes\mathbf{1} + \mathbf{1}\otimes\tilde{\mathbf{H}}) \oplus \ldots .
\end{align}
Here, $\tilde{\mathbf{H}}_{\mathrm{as}}$ is the asymptotic Hamiltonian on the Fock space, constructed from the action of $\tilde{\mathbf{H}}$ on the single-particle subspace. The operator $\mathbf{Q}$ acts on the Fock space as $\mathbf{Q}\oplus(\mathbf{Q}\otimes\mathbf{Q})\oplus\ldots\,$, so the ghost parity of a multiparticle state is the product of the individual ghost parities. It follows that $[\mathbf{Q},\tilde{\mathbf{H}}_{\mathrm{as}}]=0$ and $\tilde{\mathbf{H}}_{\mathrm{as}}^\dagger\mathbf{K}=\mathbf{K}\tilde{\mathbf{H}}_{\mathrm{as}}$. The $S$-matrix is therefore unitary with respect to either inner product:
\begin{equation}
\tilde{\mathbf{S}}^\dagger\mathbf{K}\tilde{\mathbf{S}} = \mathbf{K}. \label{e11}
\end{equation}

Since $\mathbf{Q}$ commutes with both $\tilde{\mathbf{H}}$ and $\tilde{\mathbf{H}}_{\mathrm{as}}$, it commutes with $\tilde{\mathbf{S}}$. For initial and final states of definite ghost parity,
\begin{equation}
0 = \boldsymbol{\psi}^{f\dagger}\mathbf{K}[\mathbf{Q},\tilde{\mathbf{S}}]\boldsymbol{\psi}^i = (\sigma_f - \sigma_i)\,\boldsymbol{\psi}^{f\dagger}\mathbf{K}\tilde{\mathbf{S}}\boldsymbol{\psi}^i,
\end{equation}
where we have used $\mathbf{Q}^\dagger\mathbf{K}=\mathbf{K}\mathbf{Q}$. Transition amplitudes therefore vanish between Fock states of different ghost parity. In the continuum, where in- and out-states are degenerate in energy, the absence of on-shell cross-sector transitions is part of the definition of the superselected theory. This is implemented in the perturbation theory introduced in Section~\ref{s5} and developed further in \cite{pertur}. The negative-probability problem, however, concerns not only such transition amplitudes but also whether arbitrary superpositions of states with opposite norm signs are allowed. The superselection rule excludes these superpositions by restricting the physical state space to a disjoint union of state spaces associated with the $Q = +1$ and $Q = -1$ sectors.

Writing $\tilde{\mathbf{S}}=\mathbf{1}+i\tilde{\mathbf{T}}$, the unitarity condition (\ref{e11}) becomes
\begin{equation}
-i\bigl(\mathbf{K}\tilde{\mathbf{T}} - \tilde{\mathbf{T}}^\dagger\mathbf{K}\bigr) = \tilde{\mathbf{T}}^\dagger\mathbf{K}\tilde{\mathbf{T}}.
\end{equation}
Taking the normalized diagonal matrix element for an initial state $\boldsymbol{\psi}^i$ gives
\begin{equation}
\frac{2\,\mathrm{Im}\,\boldsymbol{\psi}^{i\dagger}\mathbf{K}\tilde{\mathbf{T}}\boldsymbol{\psi}^i}{\boldsymbol{\psi}^{i\dagger}\mathbf{K}\boldsymbol{\psi}^i}
= \frac{\boldsymbol{\psi}^{i\dagger}\tilde{\mathbf{T}}^\dagger\mathbf{K}\tilde{\mathbf{T}}\boldsymbol{\psi}^i}{\boldsymbol{\psi}^{i\dagger}\mathbf{K}\boldsymbol{\psi}^i}.
\end{equation}
Inserting the completeness relation (\ref{e3}) between $\mathbf{K}$ and $\tilde{\mathbf{T}}$ on the right-hand side yields the generalized optical theorem,
\begin{equation}
\frac{2\,\mathrm{Im}\,\boldsymbol{\psi}^{i\dagger}\mathbf{K}\tilde{\mathbf{T}}\boldsymbol{\psi}^i}{\boldsymbol{\psi}^{i\dagger}\mathbf{K}\boldsymbol{\psi}^i}
= \sum_X\int d\Pi_X\,\frac{\bigl|\boldsymbol{\psi}^{X\dagger}\mathbf{K}\tilde{\mathbf{T}}\boldsymbol{\psi}^i\bigr|^2}{\boldsymbol{\psi}^{i\dagger}\mathbf{K}\boldsymbol{\psi}^i\;\boldsymbol{\psi}^{X\dagger}\mathbf{K}\boldsymbol{\psi}^X}. \label{e7}
\end{equation}

For the original ghost theory with $\mathbf{K}=\bm\eta$, this is the standard statement of unitarity. The left-hand side is calculated using $\eta$-perturbation theory and is therefore sensitive to opposite-sign ghost propagators. Each such sign is matched by a sign from the norm factor $\boldsymbol{\psi}^{X\dagger}\bm\eta\boldsymbol{\psi}^X$ on the right-hand side, a fact overlooked in claims that ``ghosts violate unitarity''. The problem is not unitarity itself, but the minus signs on the right-hand side that obstruct a probabilistic interpretation.

The Fock states in (\ref{e7}) have definite ghost parity and satisfy $\boldsymbol{\psi}^{Y\dagger}\mathbf{G}=\sigma_Y\boldsymbol{\psi}^{Y\dagger}\bm\eta$. Hence, replacing $\bm\eta$ with $\mathbf{G}$ makes all the signs on the right-hand side positive. The difficulty is that perturbation theory computes the left-hand side with respect to the $\eta$ inner product, and there is no corresponding $G$-perturbation theory. Without the superselection rule, the $\eta$ and $G$ forms of the right-hand side differ. The superselection rule reconciles them by restricting the sum over intermediate states to those that preserve $Q$. Contributions that would carry minus signs, which correspond to cross-sector transitions, vanish identically. The two formulations of the right-hand side then agree term by term on the sector-preserving physical content.

The structure of the right-hand side can be clarified further by noting that
\begin{equation}
\bigl|\boldsymbol{\psi}^{X\dagger}\mathbf{K}\tilde{\mathbf{T}}\boldsymbol{\psi}^i\bigr|^2
= \boldsymbol{\psi}^{i\dagger}\mathbf{K}\tilde{\mathbf{T}}^\#\boldsymbol{\psi}^X\boldsymbol{\psi}^{X\dagger}\mathbf{K}\tilde{\mathbf{T}}\boldsymbol{\psi}^i, \label{e6}
\end{equation}
where $\tilde{\mathbf{T}}^\# = \mathbf{K}^{-1}\tilde{\mathbf{T}}^\dagger\mathbf{K}$ is the $K$-adjoint of $\tilde{\mathbf{T}}$. In practice, $\tilde{\mathbf{T}}^\#$ is obtained from $\tilde{\mathbf{T}}^\dagger$ by replacing every occurrence of $\tilde{\mathbf{H}}^\dagger$ with $\tilde{\mathbf{H}}$. This relation gives a more precise meaning to the conjugate amplitude.

Realizing the optical theorem in a superselected ghost theory requires a sector-preserving perturbation theory, as in Section~\ref{s5} and \cite{pertur}.

\section{Spectral representation}\label{s4}

We now turn to the spectral representation of propagators. We first construct the propagator for both choices of inner product $\mathbf{K}=\bm\eta$ and $\mathbf{K}=\mathbf{G}$. We then derive its spectral decomposition and discuss how the superselection rule controls its analytic structure. In matrix notation, composing operators requires inserting the inverse metric between adjacent operator matrices. Thus, the field product $\phi(x)\phi(y)$ is represented by $\bm\phi(x)\mathbf{K}^{-1}\bm\phi(y)$. For the two-point function,
\begin{align}
\langle\bar{0}|T\phi(x)\phi(y)|\bar{0}\rangle_K &= \bm\psi^{\bar{0}\dagger} T[\bm\phi(x)\mathbf{K}^{-1}\bm\phi(y)]\bm\psi^{\bar{0}} \\
&= (\mathbf{K}\bm\psi^{\bar{0}})^\dagger T[\bm\phi_K(x)\bm\phi_K(y)]\bm\psi^{\bar{0}}, \quad \bm\phi_K \equiv \mathbf{K}^{-1}\bm\phi. \label{e5}
\end{align}
The exact vacuum is represented by the right eigenvector $\bm\psi^{\bar{0}}$ of $\tilde{\mathbf{H}}$ and the corresponding left eigenvector $(\mathbf{K}\bm\psi^{\bar{0}})^\dagger$. Since $\bm\phi^\dagger = \bm\phi$, the field $\bm\phi_K$ is self-adjoint with respect to $\mathbf{K}$: $\bm\phi_K^\dagger \mathbf{K} = \mathbf{K}\bm\phi_K$. Of the two propagators, the $\eta$-propagator is intrinsic to perturbation theory. The extension to higher-point functions is straightforward and requires additional insertions of $\mathbf{K}^{-1}$.

Consider the pseudo-Hermitian four-momentum operator $\tilde{\mathbf{P}}^\mu = \bm\eta \mathbf{P}^\mu$. It is constructed analogously to the Hamiltonian and commutes with $\mathbf{Q}$. Since $\tilde{\mathbf{P}}^\mu$ is self-adjoint with respect to $\mathbf{K}$,
$\tilde{\mathbf{P}}^{\mu\dagger}\mathbf{K} = \mathbf{K}\tilde{\mathbf{P}}^\mu$,
the spacetime translation of $\bm\phi_K$ takes the usual Heisenberg form,
\begin{equation}
\bm\phi_K(x) = e^{i\tilde{\mathbf{P}}\cdot x}\bm\phi_K(0)e^{-i\tilde{\mathbf{P}}\cdot x}. \label{e4}
\end{equation}
The factors $\bm\psi^{X\dagger}\mathbf{K}$ and $\bm\psi^X$ in the completeness relation (\ref{e3}) are the left and right eigenvectors of $\tilde{\mathbf{P}}^\mu$. Inserting (\ref{e3}) between the fields in (\ref{e5}) and using (\ref{e4}) replaces each occurrence of $\tilde{\mathbf{P}}^\mu$ with $p_X^\mu$. Using the self-adjointness of $\bm\phi_K$, the remaining field-dependent factor can be written as $|\bm\psi^{X\dagger}\mathbf{K}\bm\phi_K(0)\bm\psi^{\bar{0}}|^2$. This equals $|\bm\psi^{X\dagger}\bm\phi(0)\bm\psi^{\bar{0}}|^2 = |\langle X|\phi(0)|\bar{0}\rangle|^2$ and is therefore independent of the inner product. The inner product appears only in the norm factor $\bm\psi^{X\dagger}\mathbf{K}\bm\psi^X = \langle X|X\rangle_K$ in (\ref{e3}). The remainder of the spectral representation derivation follows in the standard way.

We now impose the superselection rule. The free fields need not be aligned with the eigenstates of $\mathbf{Q}$. When they are not, we decompose the field $\bm\phi(0)$ into two parts: $\bm\phi^+=\frac{1}{2}(\bm\phi+\mathbf{Q}^{\dagger}\bm\phi\mathbf{Q})$ creates states from the vacuum in the $Q=+1$ sector, while $\bm\phi^-=\frac{1}{2}(\bm\phi-\mathbf{Q}^{\dagger}\bm\phi\mathbf{Q})$ creates states in the $Q=-1$ sector. We label these states by $X^+$ and $X^-$, respectively. The spectral representation then decomposes into two sectorwise sums over $\{X^+\}$ and $\{X^-\}$:
\begin{align}
\langle\bar{0}|T\phi(x)\phi(y)|\bar{0}\rangle_K &= \int_0^\infty ds\,(\rho_K^+(s) + \rho_K^-(s))\,D_F(x-y,s),\\
\rho_K^+(s) &= \sum_{\{X^+\}} \delta(s - p_X^2)\,\frac{|\langle X|\phi(0)|\bar{0}\rangle|^2}{\langle X|X\rangle_K},\\
\rho_K^-(s) &= \sum_{\{X^-\}} \delta(s - p_X^2)\,\frac{|\langle X|\phi(0)|\bar{0}\rangle|^2}{\langle X|X\rangle_K},
\end{align}
where $D_F$ is the Feynman propagator. The $\eta$ and $G$ spectral representations differ only by the sign of $\langle X|X\rangle_K$ in the negative ghost parity sector, so $\rho_G^-(s) = -\rho_\eta^-(s) > 0$. We now focus on the $\eta$-propagator.

The full propagator is therefore the sum of two propagators, each receiving contributions only from states with the corresponding ghost parity and describing propagation in one superselection sector. Each sector has a K\"all\'en--Lehmann representation with spectral support for real $s$, and each sectoral propagator is analytic on the physical sheet except for real poles and cuts. This structure follows because $\rho_\eta^+(s)$ and $\rho_\eta^-(s)$ receive only positive and negative contributions, respectively. For example, a ghost pole can move onto the second sheet when a single-particle ghost state mixes with ghost multiparticle states. It cannot mix with non-ghost multiparticle states because they belong to the other sector. Thus, complex-conjugate poles cannot occur on the physical sheet of the exact $Q$-superselected theory. The approach of the next section realizes this structure in a sector-preserving perturbation theory.

In ordinary perturbation theory, complex-conjugate poles arise at finite order. Their implications for the analytic structure of ghost theories have been studied extensively \cite{Grinstein:2007mp,Anselmi:2018kgz,Donoghue:2019ecz,Salvio:2018crh}, with the problem clearly stated in~\cite{Kubo:2024ysu,Buoninfante:2025klm} and a ghost-masking effect studied in \cite{Buoninfante:2026mve,Buoninfante:2026jfj}.

The matrix fields that define the $\eta$-propagator, $\bm\phi_\eta = \bm\eta\bm\phi = \tilde{\bm\phi}$, are the same fields that appear in the matrix Hamiltonian $\tilde{\mathbf{H}}$, whose terms are products of $\tilde{\bm\phi}$ or $\tilde{\boldsymbol{\pi}}$ \cite{Holdom:2024onr}. These fields satisfy the canonical commutation relation $[\tilde{\bm\phi}(\mathbf{x}), \tilde{\boldsymbol{\pi}}(\mathbf{y})] = i\delta(\mathbf{x}-\mathbf{y})\mathbf{1}$. Their equation of motion is $\dot{\tilde{\bm\phi}}(\mathbf{x}) = i[\tilde{\mathbf{H}}, \tilde{\bm\phi}(\mathbf{x})] = \pm\tilde{\boldsymbol{\pi}}(\mathbf{x})$. The sign determines whether the field is a ghost, that is, whether the $\tilde{\boldsymbol{\pi}}^2$ term in $\tilde{\mathbf{H}}$ has the wrong sign. In either case, the free spectrum is positive \cite{Holdom:2024onr}. By the standard argument, the equal-time commutator $[\tilde{\bm\phi}(\mathbf{x}), \dot{\tilde{\bm\phi}}(\mathbf{y})]$ enters the derivation of the sum rule, and the sign $\pm$ in $\dot{\tilde{\bm\phi}}$ carries over directly:
\begin{equation}
\int_0^\infty (\rho_\eta^+(s) + \rho_\eta^-(s))\,ds = \pm 1.
\end{equation}
The sign determines which spectral function dominates the sum, and equivalently, whether the fundamental field is a ghost. In a standard QFT, the terms in the sum rule can be interpreted as probabilities, but this interpretation does not apply in superselected ghost theory. The sum rule instead relates two sets of transitions, each confined to one sector. This sectorwise structure ensures that every physical transition has a positive probability.

The $G$-propagator has a non-negative spectral function,
\begin{equation}
\int_0^\infty (\rho_G^+(s) + \rho_G^-(s))\,ds > 0,
\end{equation}
but the integral is not unity because the fields $\bm\phi_G(x)$ that define the $G$-propagator do not satisfy the canonical commutation relation.

When $X$ is a single stable particle labeled by $a$, its contribution to $\rho_\eta^+$ or $\rho_\eta^-$ produces a simple pole in the $\eta$-propagator:
\begin{equation}
\Delta(p)\big|_{p^2\to m_a^2} \simeq \frac{Z_a}{\langle a|a\rangle_\eta}\,\frac{i}{p^2 - m_a^2 + i\epsilon},
\end{equation}
where $Z_a = |\langle a|\phi(0)|\bar{0}\rangle|^2$. The norm factor determines the sign of the pole residue. The same combination of factors appears on the right-hand side of the optical theorem (\ref{e7}) when the state $X$ contains a particle $a$. Applying LSZ reduction to express the $S$-matrix element and its conjugate in terms of amputated amplitudes introduces the factors $\langle\bar{0}|\phi(0)|a\rangle$ and $\langle a|\phi(0)|\bar{0}\rangle$, respectively. The norm factor in (\ref{e7}) provides $1/\langle a|a\rangle_\eta$. Once a sector-preserving perturbation theory is available, the cutting rules for the left-hand side of (\ref{e7}) can be formulated in the standard way because the corresponding pole-residue factor is reproduced on the right-hand side.

\section{Sector-preserving perturbation theory}\label{s5}

For a perturbative calculation of $S$-matrix elements, the asymptotic Hamiltonian in the formal definition (\ref{e10}) is replaced by the free Hamiltonian $\tilde{\mathbf{H}}_0$. This replacement produces the standard Dyson series in terms of the interaction Hamiltonian $\tilde{\mathbf{H}}_1$. At finite order, however, this perturbation theory does not respect the superselection rule because neither $\tilde{\mathbf{H}}_0$ nor $\tilde{\mathbf{H}}_1$ commutes with the exact ghost parity $\mathbf{Q}$. Intermediate states of both ghost parities therefore appear. A perturbative framework that respects the exact ghost parity is essential because, thus far, the superselected theory has only a nonperturbative definition in terms of the exact ghost parity operator $Q$.

This framework can be obtained by transforming the original perturbation theory, with the positive-definite inner product $G$ playing a central role. Although the original perturbation theory is based on the $\eta$ inner product, the operator $\mathbf{g} = \mathbf{G}^{1/2}$ defines the similarity transformation
\begin{equation}
\mathbf{h} = \mathbf{g}\tilde{\mathbf{H}}\mathbf{g}^{-1}, \qquad \bm\eta = \mathbf{g}\mathbf{Q}\mathbf{g}^{-1}.
\label{e12}\end{equation}
The transformed Hamiltonian $\mathbf{h}$ is Hermitian, $\mathbf{h}^\dagger = \mathbf{h}$ \cite{Mostafazadeh:2001jk,Mostafazadeh:2001nr,Mostafazadeh:2002id}. To establish the second identity, we begin with $\mathbf{G} = \bm\eta\mathbf{Q}$. Multiplication on the right by $\mathbf{Q}\mathbf{G}$ gives $\mathbf{G}\mathbf{Q}\mathbf{G} = \mathbf{Q}$, or equivalently $\mathbf{G}\mathbf{Q}=\mathbf{Q}\mathbf{G}^{-1}$. Extending this result to $\mathbf{G}^n\mathbf{Q}=\mathbf{Q}\mathbf{G}^{-n}$ and setting $n=1/2$ yields $\bm\eta = \mathbf{G}\mathbf{Q} = \mathbf{g}^2\mathbf{Q} = \mathbf{g}\mathbf{Q}\mathbf{g}^{-1}$.

The similarity transformation maps $[\mathbf{Q}, \tilde{\mathbf{H}}] = 0$ to $[\bm\eta, \mathbf{h}] = 0$. In the transformed picture, $\mathbf{h}$ is therefore block diagonal in the sectors of positive and negative $\eta$ ghost parity. Since $\mathbf{g} = \mathbf{1}$ at zero coupling, $\mathbf{h}_0 = \tilde{\mathbf{H}}_0$. Each term $\mathbf{h}_i$ at order $i$ in the coupling expansion of $\mathbf{h}$ then also satisfies $[\bm\eta, \mathbf{h}_i] = 0$. The similarity-transformed perturbation theory thus preserves $\eta$ ghost parity order by order. Ordinary perturbation theory expands in $\eta$ eigenstates that are not aligned with the exact $Q$ eigenstates. The transformed theory also expands in $\eta$ eigenstates, but in this picture $\eta$ additionally defines the exact ghost parity and the superselection rule.

For initial and final states in a single sector, such as the $\eta=1$ sector, the transformed perturbation theory contains no cuts through opposite-sector intermediate states. Heavy ghost degrees of freedom may still affect low-energy effective interactions virtually, but they cannot appear as negative-norm physical intermediate states. This result is in accordance with standard positivity arguments for effective field theories. Above the ghost threshold, the lightest $\eta = -1$ excitation is a stable asymptotic particle in a separate superselection sector. Heavier excitations in that sector can still become resonances by coupling to same-sector multiparticle states. Pairs of stable ghost particles can be created in the $\eta=1$ sector.

To calculate $\mathbf{h}_1$, write $\tilde{\mathbf{H}} = \tilde{\mathbf{H}}_0 + \tilde{\mathbf{H}}_1$ and $\mathbf{Q} = \bm\eta + \mathbf{Q}_1+\ldots$, where $\tilde{\mathbf{H}}_1$ and $\mathbf{Q}_1$ are first order in the coupling. The relation $[\mathbf{Q},\tilde{\mathbf{H}}] = 0$ implies $[\mathbf{Q}_1, \tilde{\mathbf{H}}_0] = -[\bm\eta, \tilde{\mathbf{H}}_1]$, while $\mathbf{Q}^2 = \mathbf{1}$ implies $\bm\eta\mathbf{Q}_1 + \mathbf{Q}_1\bm\eta = 0$. Expanding $\mathbf{g}$ and $\mathbf{g}^{-1}$ to first order then gives
\begin{equation}
   \mathbf{h}_1 = \frac{1}{2}(\tilde{\mathbf{H}}_1 + \bm\eta\tilde{\mathbf{H}}_1\bm\eta).
\end{equation}
This is the projection of $\tilde{\mathbf{H}}_1$ onto its sector-preserving part. Thus, at lowest order, the transformed interaction is the original local interaction with all ghost-parity-changing matrix elements projected out. This $\mathbf{h}_1$ contains the local couplings of the transformed theory.

At higher orders in the coupling, the new effective vertices $\mathbf{h}_i$ are nonlocal and progressively more complicated \cite{pertur}, but every vertex preserves $\eta$ ghost parity. This construction provides a perturbative completion of the optical theorem and the spectral representation discussed in the previous sections. Although the original perturbation theory fails to achieve this, in \cite{pertur} we discuss the sense in which the renormalization and the UV divergence structure of the two theories agree with each other.

The original ghost perturbation theory may also remain useful for describing hard scattering at high energies, such as ultra-Planckian scattering \cite{Holdom:2021hlo} in quantum quadratic gravity (QQG) \cite{Stelle:1976gc,Salvio:2018crh}. This is motivated by the analogy to the calculation of high-energy scattering cross sections in QCD. Perturbative QCD does not make confinement manifest, in particular $Q^{a}|\psi\rangle=0$ where $Q^{a}$ are the generators of global color rotations, any more than the original ghost perturbation theory makes manifest the superselection rule, in particular $Q|\psi\rangle=\pm|\psi\rangle$. Nevertheless, perturbative QCD is used for high-energy scattering. This is related to how parton-shower phenomena in the initial and final states imply that the hard scattering process occurs far off shell. Inclusive hard cross sections are then of interest, and these can be obtained from perturbative QCD because the on-shell quarks and gluons provide a complete set of states.

Similarly, in QQG one may be able to ignore the superselection structure to calculate the hard scattering cross section inclusively among ghosts and non-ghosts in the original ghost perturbation theory. Cancelations in the sum then produce sensible high-energy behavior \cite{Holdom:2021hlo}. It is the transition between the actual initial state, before parton showers, and the actual final state, after parton showers, that must obey the superselection rule. This transition is described by the sector-preserving perturbation theory, but when it comes to the virtual high energy scattering process, there appears to be a shortcut.

\section{Conclusion}\label{s6}

We have proposed promoting exact ghost parity $Q$ to a superselection charge in the real-spectrum regime of an interacting ghost theory. The operator $Q$ is determined by the spectral decomposition of the Hamiltonian, reduces to free ghost parity at zero coupling, and assigns to each energy eigenstate the sign of its native norm. Treating $Q$ as a superselection charge is an additional physical postulate that defines a new theory, the superselected ghost theory. Its physical states have definite ghost parity and its observables commute with $Q$.

This postulate gives the native indefinite inner product a consistent probabilistic role. Because physical superpositions and observables preserve the $Q$ sectors, the native Born rule yields non-negative probabilities. The optical theorem formulated with the native inner product similarly has a probability interpretation. Superselection also decomposes the propagator into sectorwise spectral representations with fixed-sign spectral functions. Each sectoral propagator has the standard analytic structure on the physical sheet, with no complex-conjugate poles.

Ordinary perturbation theory does not preserve exact ghost parity at finite order. The similarity transformation introduced here maps the full Hamiltonian to a Hermitian Hamiltonian that is block diagonal in free ghost parity. The resulting perturbation theory preserves the superselection sectors order by order and provides a perturbative realization of the optical theorem and spectral representation. Its higher-order interactions are generally nonlocal and result in a modified form of old-fashioned perturbation theory \cite{pertur}.

The nonlocality is central to this construction; because $Q$ is determined by the entire interacting spectrum, it is generally nonlocal. Consequently, using $Q$ to project a local operator onto its $Q$-even part can make the corresponding physical observable nonlocal. The same nonlocality reappears in the higher-order interactions of the transformed perturbation theory. This structure is particularly suggestive for gravity, where general covariance forbids local gauge-invariant observables and shifts physical content toward asymptotic data. The ghost in renormalizable quantum quadratic gravity may be a similar signal that the gravitational sector replaces local fields in the interaction region with an asymptotic $S$-matrix description.

Matter fields of the Standard Model are expected to commute with $Q$ and so their usual local observables remain unaffected. The lightest gravitational excitation with $Q=-1$ is kinematically stable under exact superselection and may therefore be a dark-matter candidate, although its phenomenology remains to be established. While the effective theory below the lightest ghost is ghost free, the underlying superselected ghost theory will leave an imprint on its effective interactions.

The superselected theory has a classical limit, but it is not that of the classical four-derivative theory. It is defined by the tree-level part of the sector-preserving perturbation theory \cite{pertur}. The resulting classical response to non-ghost sources is standard only at energies below the lightest ghost mass.  The difference lies on the ghost mass shell, where the negative-energy radiation of the classical four-derivative theory is absent, as is single-ghost radiation; ghosts are instead radiated in pairs with positive energy. A classical field requires many quanta, and states of even and odd ghost number are then distinguished only by observables sensitive to the ghost parity itself, and not by local observables such as the energy density. In this way, both sectors are physical and they both have a classical limit.

The principal open question is dynamical: whether the real-spectrum regime occurs in interacting 3+1D theories of practical interest, especially quantum quadratic gravity. The complementary regime containing complex-conjugate energy pairs requires an additional $\mathbb Z_2$ structure and is considered separately \cite{entang}. The present paper presents a self-consistent real-spectrum regime within this broader picture.

The indefinite inner product is not merely a pathology to be removed. Combined with exact ghost-parity superselection, it becomes the structure that allows a unitary ghost theory to support probabilities and asymptotic observables.\\

\emph{Declaration of generative AI and AI-assisted technologies in the manuscript preparation process:}
Various LLMs were guided by the author to comment on the content and prose of various versions of this manuscript. The author reviewed and edited all suggestions and takes full responsibility for the final version.

\end{document}